\documentclass[a4paper,conference]{IEEEtran}

\usepackage{fancyhead}
\usepackage{psfrag}
\usepackage{epsfig}
\usepackage{graphicx}
\usepackage{amsmath}
\usepackage{amssymb}
\usepackage{amsthm}
\usepackage{subfigure}
\usepackage{cite}
\usepackage[utf8]{inputenc}
\usepackage{color}
\usepackage{url}
\usepackage{tikz}
\usepackage{pgfplots}
\usepackage[english]{babel}
\usepackage{algorithm}
\usepackage{algorithmic}
\usepackage{trfsigns}
\usepackage{romannum}
\usepackage{relsize}

\newcommand{\fra}{\mathcal{A}}
\newcommand{\frb}{\mathcal{B}}
\newcommand{\frr}{\mathcal{RM}}
\newcommand{\C}{\mathcal{C}}
\newcommand{\FF}{\mathbb{F}}
\newcommand{\Fq}{\mathbb{F}_q}
\newcommand{\myDFT}{\ensuremath{\mathcal{F}}}
\newcommand{\myiDFT}{\ensuremath{\mathcal{F}^{-1}}}
\newcommand{\RS}[2]{\ensuremath{\mathcal{RS}(q;#1,#2)}}
\newcommand{\RSq}[1]{\ensuremath{\mathcal{RS}(#1)}}

\newcommand{\PR}[1]{\ensuremath{\Prob\left(#1\right)}}
\DeclareMathOperator{\dH}{d_H}
\DeclareMathOperator{\wtH}{wt_H}
\DeclareMathOperator{\Prob}{P}
\newcommand{\ve}[1]{\ensuremath{\mathbf{#1}}}
\renewcommand{\aa}{\ve{a}}
\newcommand{\bb}{\ve{b}}
\newcommand{\ee}{\ve{e}}
\newcommand{\yy}{\ve{y}}

\newcommand{\Romannumcolor}[1]{\ensuremath{\textcolor{blue}{\Romannum{#1}}}}
\definecolor{darkgreen}{rgb}{0,0.3,0}
\newcommand{\Step}[1]{\ensuremath{\textcolor{red}{(#1)}}}

\IEEEoverridecommandlockouts

\newtheorem{definition}{Definition}

\begin{document}

\title{On Error Correction for \\Physical Unclonable Functions}

\author{\IEEEauthorblockN{Sven Puchinger$^{\ast}$, Sven Müelich$^{\ast}$, Martin Bossert$^{\ast}$, Matthias Hiller$^{\diamond}$, Georg Sigl$^{\diamond}$}
\IEEEauthorblockA{$^{\ast}$Institute of Communications Engineering, University of Ulm, Germany}
\texttt{\{sven.puchinger | sven.mueelich | martin.bossert\}@uni-ulm.de}
\IEEEauthorblockA{$^{\diamond}$Institute for Security in Information Technology, Technische Universität M\"unchen, Germany}
\texttt{\{matthias.hiller | sigl\}@tum.de}
}
\maketitle

\begin{abstract}
Physical Unclonable Functions evaluate manufacturing variations to generate
secure cryptographic keys for embedded systems without secure key storage.
It is explained how methods from coding theory are applied in order to ensure reliable key reproduction.
We show how better results can be obtained using code classes and decoding principles not used for this scenario before.
These methods are exemplified by specific code constructions which improve existing codes with respect to error probability, decoding complexity and codeword length.
\end{abstract}


\vspace{-0.25cm}
\section{Introduction}
Cryptographic applications require random, unique and unpredictable keys.
Since most cryptosystems need to access the key several times, it usually has to be stored permanently, which is a potential vulnerability regarding security.
Implementing secure key generation and storage is therefore an important and challenging task.

A \emph{Physical Unclonable Function} (PUF) is a, typically digital, circuit that 
possesses an intrinsic randomness due to process variations during manufacturing and can therefore be used to generate a key.
This key can be reproduced on demand.
However, the PUF output when reproducing a key varies, which can be interpreted as errors. Thus, error correction must be used in order to compensate this effect.
Previous work on this topic used standard constructions, e.g. an ordinary concatenated scheme of a BCH and Repetition code in \cite{MaesCryptoPaper2012}.
In this paper, we extend our results from \cite{MueelichACCT2014} and propose code constructions based on \emph{generalized concatenated}, \emph{Reed--Muller} and \emph{Reed--Solomon codes} for the application with PUFs, which have advantages with respect to decoding complexity, error correction capability and code length.
The paper first describes PUFs and explains how coding theory is applied to realize key generation and reproduction using PUFs.
Section~\ref{sec_constructions} describes methods and codes suitable for this scenario.
Finally, specific code constructions, improving those commonly used for PUFs, exemplify these methods in Section~\ref{sec_examples}.
We summarize the results in the last section.

\section{Physical Unclonable Functions}
\label{sec_pufs}
In \cite{MaesDiss2012}, a PUF is described as a physical entity which uses an input (challenge) in order to produce an output (response), where a challenge can result in different responses when applied to a certain PUF instance several times.
The distance of two such responses is called \emph{intra-distance}\footnote{With distance we mean the \emph{Hamming} distance $\dH$.}.
Reasons for these varying responses are random noise, measurement uncertainties, aging and changing environmental conditions like temperature or supply voltage.
A small response intra-distance is preferred, since there is a need for reproducibility of responses.
The distance of the responses of two different PUF instances using the same challenge is called \emph{inter-distance}, and results from variations during the manufacturing process.
This measure gives us the distinguishability of different PUF instances, which is preferred to be large.
Unclonable means the hardness of manufacturing two PUFs with the same challenge-response-behavior.
There are many possibilities to realize PUFs, e.g. delay-based (e.g. Ring Oscillator PUFs) or memory-based (e.g. SRAM PUFs).
An overview of popular types can be found in \cite{MaesDiss2012}.

PUFs can be used in order to realize secure key generation and storage for cryptographic applications.
Due to static randomness over the PUFs lifetime, it is possible to regenerate a key repeatedly on demand instead of storing it permanently.
As described above, PUF responses are not exactly reproducible and therefore a response cannot be used as key directly.
Hence, methods of coding theory must be used.

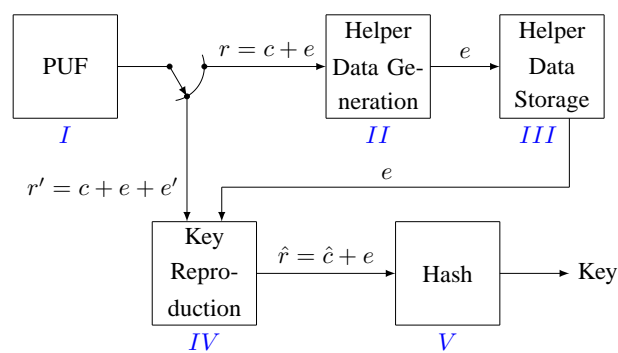
\begin{figure}[h]
{
\resizebox{0.45\textwidth}{!}{
\centering
\begin{tikzpicture}[scale=0.5]

\draw (0,0) -- (0,3);
\draw (0,3) -- (3,3);
\draw (3,3) -- (3,0);
\draw (3,0) -- (0,0);
\draw (1.5,1.5) node {PUF};
\draw (1.5,0) node [below] {$\Romannumcolor{1}$};

\draw (3,1.5) -- (4.5,1.5);
\draw[->,>=latex] (5,1.5-0.86602540378) -- (5,-3);
\draw[fill] (4.5,1.5) circle (2pt);
\draw[fill] (5.5,1.5) circle (2pt);
\draw[fill] (5,1.5-0.86602540378) circle (2pt);
\draw (5,-2) node [left] {$r' = c + e + e'$};
\draw [domain=-80:20] plot ({4.5+cos(\x)}, {1.5+sin(\x)});
\draw[->,>=latex] (4.5,1.5) -- (5,1.5-0.86602540378);

\draw (9,0) -- (9,3);
\draw (9,3) -- (12,3);
\draw (12,3) -- (12,0);
\draw (12,0) -- (9,0);
\draw (10.5,2.5) node {Helper};
\draw (10.5,1.5) node {Data Ge-};
\draw (10.5,0.5) node {neration};
\draw (10.5,0) node [below] {$\Romannumcolor{2}$};

\draw[->,>=latex] (5.5,1.5) -- (9,1.5);
\draw[->,>=latex] (12,1.5) -- (14,1.5);
\draw (7.3,1.5) node [above] {$r = c + e$};
\draw (13,1.5) node [above] {$e$};

\draw (14,0) -- (14,3);
\draw (14,3) -- (17,3);
\draw (17,3) -- (17,0);
\draw (17,0) -- (14,0);
\draw (15.5,2.5) node {Helper};
\draw (15.5,1.5) node {Data};
\draw (15.5,0.5) node {Storage};
\draw (15.1,0) node [below] {$\Romannumcolor{3}$};

\draw (16,0) -- (16,-2);
\draw (16,-2) -- (6,-2);
\draw[->,>=latex] (6,-2) -- (6,-3);
\draw (10.75,-2) node [above] {$e$};

\draw (4,-3) -- (7,-3);
\draw (7,-3) -- (7,-6);
\draw (7,-6) -- (4,-6);
\draw (4,-6) -- (4,-3);
\draw (5.5,-3.5) node {Key};
\draw (5.5,-4.5) node {Repro-};
\draw (5.5,-5.5) node {duction};
\draw (5.5,-6) node [below] {$\Romannumcolor{4}$};

\draw[->,>=latex] (7,-4.5) -- (11,-4.5);
\draw (9,-4.5) node [above] {$\hat{r} = \hat{c} + e$};

\draw (11,-3) -- (14,-3);
\draw (14,-3) -- (14,-6);
\draw (14,-6) -- (11,-6);
\draw (11,-6) -- (11,-3);
\draw (12.5,-4.5) node {Hash};
\draw (12.5,-6) node [below] {$\Romannumcolor{5}$};

\draw[->,>=latex] (14,-4.5) -- (16,-4.5);
\draw (16,-4.5) node [right] {Key};

\end{tikzpicture}
}
}
\caption{Key Generation and Reproduction in PUFs.}
\label{fig:puf}
\end{figure}

One way to realize key reproduction is the \emph{Code-Offset Construction} \cite{Dodis2008} 
(cf. Figure~\ref{fig:puf}). 
First, for a given challenge a response $r$ is generated by the PUF (\Romannumcolor{1}). The \emph{Helper Data Generation} (\Romannumcolor{2}) subtracts a random codeword $c$ of a given code $\mathcal{C}(n,k,d)$ from $r$ and stores the result $e=r-c$ in the \emph{Helper Data Storage} (\Romannumcolor{3}). Afterwards, the response $r$ can be deleted. Hence, if an attacker is able to read this storage, he is left with an uncertainty as large as the number of codewords.

For regenerating the original response, the same challenge must be used to obtain a response $r'$ which is likely to differ slightly from $r$. For most PUFs, this can be interpreted as an additive error $r'=r+e'=c+e+e'$, resulting from a binary symmetric channel (BSC) with crossover probability $p$, where $p$ is given by the PUF. If $\wtH\left(e'\right)=\dH\left(r,r'\right)$ is within the error correction capabilities of the code, the \emph{Key Reproduction} (\Romannumcolor{4}) procedure is able to reproduce the first-time response $r$ by decoding $r'-e=c+e+e'-e =c+e'$ with a decoder of the code $\mathcal{C}(n,k,d)$ (decoding result $\hat{c}$).

Other possibilities for implementing key reproduction are the \emph{Syndrome Construction} \cite{Dodis2008}, \emph{Index-Based Syndrome Coding} \cite{Yu2010}, \emph{Complementary Index-Based Syndrome Coding} \cite{Hiller2012} and \emph{Differential Sequence Coding} \cite{Hiller2013}. One main challenge is to find good codes that can be used for key reproduction.

If an attacker is able to read $e$ from the helper data storage, his uncertainty about the response $r$ is equal to the uncertainty of $c$, namely $2^k$ codewords.
Hence, the uncertainty of the extracted key cannot be larger than the dimension of the used code.
Also, this uncertainty is even smaller due to the fact that the PUF responses themselves are not necessarily independent and uniformly distributed.
Since we want to obtain a uniformly distributed cryptographic key, $\hat{r}$ must be hashed by a cryptographic hash function (\Romannumcolor{5}) before it can be used.
The combination of key reproduction and a hash function is usually referred to as \emph{Fuzzy Extractor} \cite{dodis2008fuzzy}.

\section{Code Constructions}
\label{sec_constructions}

Code design for key reproduction in PUFs is analog to standard problems in coding theory for a given channel, e.g. a BSC with crossover probability $p$.
A typical goal is to design a code with a block error probability $\Prob_{err}$ smaller than a certain threshold.
The dimension of the code must be at least the length of the key that should be generated.
Also, the designed codes must be binary.
The length of the codewords can be chosen arbitrarily, but for generating one key, at least as many bits as the codeword length have to be extracted from the PUF, which determines the size of the PUF.
Since the decoder is usually part of an embedded security device, the decoding method must be easy to implement in hardware.
Here, we describe suitable construction and decoding methods.

\subsection{Generalized Concatenated Codes}
\label{subsec_gcc}

The authors of \cite{Bosch2008} found that concatenated codes are advisable for implementing key reproduction.
Instead of ordinary concatenated codes, we propose using \emph{Generalized Concatenated} (GC) codes as introduced in \cite{Zyablov_Shavgulidze_Bossert_1999} and \cite{Bossert1999}.
A GC code with given $n$ and $d$ contains more codewords and hence has a higher code rate than an ordinary concatenated code with the same parameters.

The main idea of GC codes is to partition an inner code $\frb^{(1)}$ of length $n_{i}$ into multiple levels of subcodes.
Let $\frb^{(i)}_{j}$ denote the $j$-th subcode at partition level $i$.
The goal is to create partitions such that the minimum distances of the subcodes increase strictly monotonically from level to level in the partition tree. Each codeword of $\frb^{(1)}$ can be uniquely determined using a numeration of the partition. This numeration is protected by outer codes. Code $\fra^{(i)}$ of length $n_{o}$ denotes the outer code which protects the numeration of the partition from level $i$ to level $i+1$. For a detailed description of GC codes, we refer to \cite{Bossert1999}.

\subsection{Reed--Muller Codes}
\label{subsec_rm}

A \emph{Reed--Muller} (RM) code $\frr(r,m)$ of order $r$ with $r \leq m$ is a binary linear code with parameters $n = 2^{m}$, $k = \sum_{i = 0}^{r} \binom{m}{i}$ and $d=2^{m-r}$.
It can be defined recursively using the \emph{Plotkin Construction} \cite{Bossert1999}:
\begin{align*}
\frr(r,m) &:= \left\{ (\ve{a} | \ve{a} + \ve{b} ) : \begin{array}{l} \ve{a} \in \frr(r,m-1) \\ \ve{b} \in \frr(r-1,m-1) \end{array}\right\}
\end{align*}
with $\frr(0,m) := \C(2^m,1,2^m)$ (Repetition code) and $\frr(m-1,m) := \C(2^m,2^m-1,2)$ (Parity Check code) for all $m$.
$\frr(1,m)$ codes are called \emph{Simplex} codes.

RM codes work well for PUF key reproduction due to an easily implementable decoding, which can also be done recursively using Algorithm~\ref{alg:rm_decoding}, which can correct up to $\tau$ errors and $\delta$ erasures if $2\tau+\delta<d$.
Erasures are treated as third symbol $\otimes$ besides $0$ and $1$, and the operation $+$ is extended such that $\otimes + x = x + \otimes := \otimes$ for all $x \in \{0,1,\otimes\}$.
Within the description of Algorithm~\ref{alg:rm_decoding}, $+$ is applied component-wise.
Decoding of the repetition and parity check codes (base cases of the recursion) works as usual by ignoring all code positions with erasures.
Alternatively, RM codes can be defined as GC codes and decoded using the algorithm described in \cite{SB94}.
\begin{algorithm}
\caption{Recursive $\frr(r,m)$ Decoder \cite{bossert2012einfuhrung}}          
\label{alg:rm_decoding}
\small
\begin{algorithmic}[1]
    \REQUIRE $\yy = (\yy_a | \yy_b) = (\aa+\ee_a | \aa+\bb+\ee_b) \in \{0,1,\otimes\}^{2^m}$
    \STATE Decode $\yy_a + \yy_b = \bb+\ee_a+\ee_b$ in $\frr(r-1,m-1)$ $\Rightarrow$ $\ve{\hat{b}}$
    \STATE Dec. $\yy_b + \ve{\hat{b}} = \aa+(\bb+\ve{\hat{b}})+\ee_a+\ee_b$ in $\frr(r,m-1)$ $\Rightarrow$ $\ve{\hat{a}}_1$
    \STATE Dec. $\yy_a = \aa+\ee_a$ in $\frr(r,m-1)$ $\Rightarrow$ $\ve{\hat{a}}_2$
    \STATE Find $i \in \{1,2\}$ such that $\dH(\yy, (\ve{\hat{a}}_i | \ve{\hat{a}}_i+\ve{\hat{b}}))$ minimal
    \RETURN $(\ve{\hat{a}}_i | \ve{\hat{a}}_i+\ve{\hat{b}})$
\end{algorithmic}
\end{algorithm}

Using RM codes in the PUF scenario is reasonable because Algorithm~\ref{alg:rm_decoding} can be implemented efficiently.
Since it can handle both errors and erasures, it also works in combination with Generalized Minimum Distance decoding (cf. Section \ref{subsec_GMD}).
Furthermore, RM codes are proper for partitioning because $\frr(r_{i},m) \subseteq \frr(r_{j},m)$ for all $r_{i} \leq r_{j}$ and partitioning of linear block codes into cosets of a linear subcode can be done easily \cite{Bossert1999}.
This property makes them suitable in a GC code.
However, RM codes are not maximum distance separable.
Also, the dimension $k$ cannot be chosen arbitrarily.
RM codes have been used before for key reproduction in PUFs \cite{maes2009low, Hiller2012}.

\subsection{Reed--Solomon Codes}
\label{subsec_rs}

\emph{Reed--Solomon} (RS) codes are one of the most commonly used codes in applications of coding theory due to the existence of efficient decoding algorithms. We describe the basics of RS codes according to \cite{Bossert1999}. Let $\Fq$ be a finite field and $\alpha$ a generator of $\Fq^{\ast}$.

\begin{definition}[Discrete Fourier Transform (DFT)]
For a polynomial $c(x) \in \Fq[x]$ with $\deg c(x) < n$, the DFT $C(x) = \myDFT\left\{c(x)\right\} \, \Laplace \, c(x)$ is defined by
\begin{align*}
C_j = n^{-1} c(\alpha^{-j}) \quad \forall j \in \left\{ 0, \dots, n-1 \right\}
\end{align*}
and the inverse DFT $\tilde{c}(x) := \myiDFT\left\{ C(x) \right\} \, \laplace \, C(x)$ is
\begin{align*}
\tilde{c}_i = C(\alpha^i) \quad \forall i \in \left\{ 0, \dots, n-1 \right\}
\end{align*}
\end{definition}

\begin{definition}[Reed--Solomon codes]
A Reed--Solomon code over a field $\Fq$ is defined as
\begin{align*}
\RS{n}{k} = \left\{ c(x) \, \laplace \, C(x) : \deg C(x) < k \right\}
\end{align*}
\end{definition}
RS codes are \emph{maximum distance separable} (MDS), that means $d = n-k+1$.
There are several algorithms for decoding RS codes both for decoding up to half the minimum distance and beyond.
An overview of the most important decoding methods for RS codes can be found in \cite{bossert2013unified}.

\begin{figure}[h]
\newlength\figureheight
\newlength\figurewidth
\setlength\figureheight{3cm}
\setlength\figurewidth{0.4\textwidth} 
\input{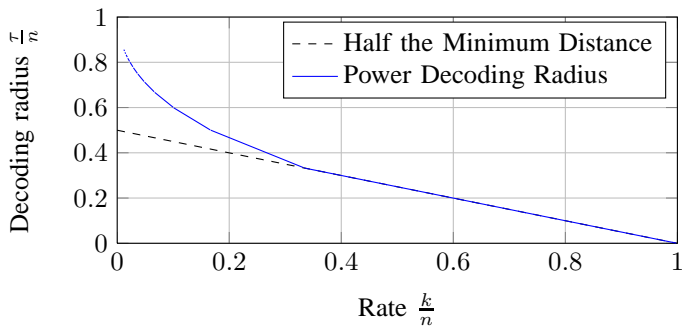}
\caption{Maximum Decoding Radius for Power Decoding RS Codes.}
\label{fig:power_decoding_radius}
\end{figure}

In this paper, we use the method of \emph{Power Decoding} \cite{schmidt2006decoding} which is easily implementable using \emph{Shift-Register Synthesis} and can correct beyond half the minimum distance for small code rates. Since we use RS codes with small rates in our construction, this method suits perfectly.
The idea of Power Decoding is to power the received word $r(x) = c(x) + e(x)$ with some positive integer $\ell$:
{\small
\begin{align*}
r^{[\ell]}(x)  :=& \sum\limits_{i=0}^{n-1} r_i^\ell x^i = \sum\limits_{i=0}^{n-1} (c_i + e_i)^\ell x^i = \sum\limits_{i=0}^{n-1} \left(\sum\limits_{j=0}^{\ell} {\ell \choose j} c_i^j e_i^{\ell-j} \right)x^i \\
				=& \sum\limits_{i=0}^{n-1} (c_i^\ell + \tilde{e}^{\ell}_i) x^i = c^{[\ell]}(x) + \tilde{e}^{[\ell]}(x)
\end{align*}}
such that for some $i$, $e_i = 0$ yields $\tilde{e}_i = 0$, but not necessarily the other way round. Hence, $\wtH(\tilde{e}^{[\ell]}(x)) \leq \wtH(e(x))$ and the indices of the nonzero coefficients of $\tilde{e}^{[\ell]}(x)$ are a subset of those of $e(x)$.
From the properties of the DFT, we know that $c^{[\ell]}(x) \, \laplace \, (C(x))^\ell$. Since $\deg C(x) \leq k-1$ implies $\deg (C(x))^\ell \leq \ell(k-1)$, we know that $c^{[\ell]}(x)$ is a codeword of $\RS{n}{k^{(\ell)} := \ell(k-1)+1}$ for all $\ell$ with $\ell(k-1)+1 \leq n$. We denote the maximum $\ell$ such that this inequality is fulfilled by $\ell_{max}$.
This approach is usually referred to as \emph{Virtual Interleaving}.
Since we know that the errors in all received words $r^{[\ell]}(x)$ are at the same positions, collaborative decoding as described in \cite{schmidt2009collaborative} can be used to improve the decoding capability.
It is shown in \cite[Section~V]{schmidt2006decoding} that, except for a negligible probability of decoding failure, Power Decoding using powers up to $\ell \leq \ell_{max}$ can correct up to
\begin{align}
\tau_{\ell} := \left\lfloor \frac{2 \ell n - \ell (\ell+1) k + \ell (\ell-1)}{2(\ell+1)} \right\rfloor \label{eq:tau_max_power}
\end{align}
errors. It can also be shown that $\ell_{max}$ is upper bounded by
\begin{align}
\ell_{max} \leq \frac{\sqrt{(k+3)^2+8(k-1)(n-1)}-(k+3)}{2(k-1)} \label{eq:l_max_power}
\end{align}
Combining \eqref{eq:tau_max_power} and \eqref{eq:l_max_power}, we obtain a maximum error correction radius as shown in Figure~\ref{fig:power_decoding_radius}. Note that for low rate codes the algorithm can correct far beyond half the minimum distance.

Most RS decoding algorithms can be modified such that they can correct erasures and errors \cite{Bossert1999}. This modification has the same effect as decoding a punctured RS code. Assuming that when transmitting a codeword from $\RS{n}{k}$, $\delta$ erasures and $\tau$ errors occurred, we can simply transform the decoding problem into correcting $\tau$ errors in $\RS{n-\delta}{k}$. So decoding is successful if $\delta+2\tau<d=n-k+1$. However, since the rate of the code used in the transformed problem is larger than the original, Power Decoding might not be helpful when a lot of erasures occur.

The main benefit of RS codes is their flexibility. The parameter $k$ (respectively $d$) can be chosen arbitrarily. Additionally, Power Decoding can be applied. Finally, RS codes have better decoding properties than RM codes. However, if the same decoding algorithm should be used for all outer codes, the codes must be defined over the same finite field $\mathbb{F}_{q^{m}}$. For the dimensions of the inner codes of two subsequent partition levels $j$ and $j-1$, it must be $k_{j-1} - k_{j} = m$, where $m = 2^{m}$ is the size of the field. Since $n_{o} \leq 2^{m}$, $m$ must be chosen sufficiently large.

\subsection{Generalized Minimum Distance Decoding}
\label{subsec_GMD}

\emph{Generalized Minimum Distance} (GMD) decoding (cf. \cite{forney1966generalized}) is a method to increase the number of correctable errors beyond half the minimum distance by incrementally declaring the least reliable positions of a received word to be erasures. Hence, \emph{soft-information} and \emph{error-erasure} decoders are needed.

\subsection{Maximum Likelihood Decoding}
\label{subsec_ML}

\emph{Maximum Likelihood} (ML) decoding finds the most likely codeword with respect to the received word. Thus, decoding only fails if two or more codewords are equally likely. For most codes, no sufficiently fast ML decoder exists, but it is applicable to codes with small dimension. Since the inner codes of GC codes often fulfill this requirement, we use ML decoders in order to decrease decoding failure probabilities of the inner codes.

\section{Code Constructions using GC Codes}
\label{sec_examples}

In this section, we show how codes for key reproduction in PUFs can be constructed using RM and RS codes in combination with GC codes.
First, we describe how codes for PUFs can be constructed based on GC codes in general.
As a starting point, the desired codeword length $n$ and a dimension $k$ which is at least the key size must be chosen.
If the information theoretic uncertainty of the source is small, a larger $k$ together with a hash function can be used to create an output with good cryptographic properties.
Next, two numbers $n_{i}$ and $n_o$ such that $n_{i}n_{o} = n$ must be found, where $n_{i}$ denotes the length of the inner codes and $n_{o}$ denotes the length of the outer codes.
$n_i$ must be chosen large enough such that an inner code with this length exists which can be partitioned easily.
The easiest way to define a partitioning is to take a linear code and a linear subcode of it.
Then, all distinct cosets of this subcode form a partitioning of the code.
If the large code has dimension $k_i$ and the subcode has dimension $k_{i+1} < k_i$, then the number of partitions is $q^{k_i-k_{i+1}}$ (here, we only consider binary inner codes, so $q=2$).
Afterwards, good outer codes with length $n_o$ have to be chosen to protect the partition indices for each partition level.
The dimensions of the codes must be chosen such that their sum is equal to the desired dimension $k$ of the entire code.
Usually, the dimensions are chosen such that they increase with the partition level.

We already described in Section~\ref{subsec_rm} why RM codes are suitable as inner codes.
Due to their easily implementable decoding algorithms, they can also be used as outer codes.
An example construction using only RM codes is given in Section~\ref{subsec:RMEx}, which has better properties than the codes commonly used for error correction in PUFs.
However their dimension cannot be chosen arbitrarily, which restricts their use as outer codes.
Therefore, in Section~\ref{subsec:RSEx} we also show how RS codes can be used instead.

\subsection{Reed-Muller Example Construction}
\label{subsec:RMEx}

In \cite{MaesCryptoPaper2012}, a design for cryptographic key generators based on PUFs was introduced, using a concatenation of a $(318,174,35)$ BCH code and a $(7,1,7)$ Repetition code in order to generate a $128$ bit key with error probability $\Prob_{err}=10^{-9}$.
The paper considers error models with BSC crossover probabilities $p$ ranging from $0.12$ to $0.14$, leading to different PUF entropies.
The higher this entropy is, the fewer bits are needed to hash to the same key size.
For a minimum code dimension, we consider the maximum entropy case with $p=0.14$.
For a fair comparison to \cite{MaesCryptoPaper2012}, the block error probability $\Prob_{err}$ should be roughly $10^{-9}$ for a $128$ bit key.
Thus, we have to choose a code with dimension $\geq 128$ and aim for a block length less than the one used in \cite{MaesCryptoPaper2012}, namely $2226$.

We give a more detailed description and analysis of the example code construction which we introduced in \cite{MueelichACCT2014}.
The example improves existing schemes in code length, block error probability and easiness of the implementation.
We choose a generalized concatenation of an inner $(16,5,8)$ Simplex code $\frb^{(1)}$ and RM codes of length $128$ as outer codes $\fra^{(i)}$.
Hence, we obtain a code of length $128 \cdot 16 = 2048$, i.e. it can be represented as a matrix with $128$ rows, each containing a codeword of the Simplex code.

\begin{figure}[h]
\tikzstyle{level 1}=[level distance=4cm, sibling distance=9cm]
\tikzstyle{level 2}=[level distance=4cm, sibling distance=3.5cm]
\tikzstyle{level 3}=[level distance=4cm, sibling distance=2.5cm]
\tikzstyle{level 4}=[level distance=4cm, sibling distance=1.5cm]

\tikzstyle{bag} = [text width=10em, text centered]
\tikzstyle{end} = [circle, minimum width=3pt,fill, inner sep=0pt]

\centering
\hspace{-1cm}
\resizebox{0.5\textwidth}{!}{
\begin{tikzpicture}[scale=0.5] 
\draw (0,-2) 	node {$\dots$};
\draw (0,-4) 	node {$\dots$};
\draw (0,-6) 	node {$\dots$};
\draw (0,-8) 	node {$\dots$};
\draw (10,0) 	node {Level 1};
\draw (10,-4) 	node {Level 2};
\draw (10,-8) 	node {Level 3};
\draw (10,-2) 	node {$\longleftarrow \fra^{(1)}(128,8,64)$};
\draw (10,-6) 	node {$\longleftarrow \fra^{(2)}(128,99,8)$};
\node[bag] {$\frb^{(1)}(16,5,8)$}
    child {
        node[bag] {$\frb^{(2)}_{0000}(16,1,16)$}       
            child {
                node[bag] {$\frb^{(3)}_{0000,0}$}
                edge from parent
                node[left] {$0$}
            }
            child {
                node[bag] {$\frb^{(3)}_{0000,1}$}
                edge from parent
                node[right] {$1$}
            }
            edge from parent
            node[left] {$0000$}
    }
    child {
        node[bag] {$\frb^{(2)}_{1111}(16,1,16)$}
            child {
                node[bag] {$\frb^{(3)}_{1111,0}$}
                edge from parent
                node[left] {$0$}
            }
            child {
                node[bag] {$\frb^{(3)}_{1111,1}$}
                edge from parent
                node[right] {$1$}
            }
            edge from parent
            node[right] {$1111$}
    };
\end{tikzpicture}
}
\caption{Partition of the inner code $\frb^{(1)}(16,5,8)$.}
\label{fig:partition_tree}
\end{figure}
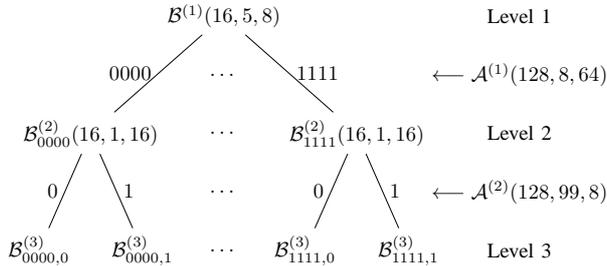

\begin{figure}[h]
\begin{flushright}
\resizebox{0.45\textwidth}{!}{
\begin{tikzpicture}[scale=0.6]

\draw [thick] (0,4) -- (0,6);
\draw [thick] (0,6) -- (1,6);
\draw [thick] (1,6) -- (1,4);
\draw [thick] (1,4) -- (0,4);
\draw (0.5,4) node[below] {$\Romannumcolor{1}$};
\draw (0,5) node[left] {8};
\draw (0.5,6) node[above] {4};

\draw[->, ultra thick] (1,5) -- (6,5);
\draw (1.75,5) node[below] {$\Step{a}$};

\draw [thick] (2.5,6) -- (3,6);
\draw [thick] (3,6) -- (3,1);
\draw [thick] (3,1) -- (2.5,1);
\draw [thick] (2.5,1) -- (2.5,6);
\draw (2.75,1) node[below] {$\Romannumcolor{2}$};
\draw (2.5,3.5) node[left] {99};
\draw (2.75,6) node[above] {1};

\draw[->, ultra thick] (3,2) -- (9,2);
\draw (4.5,2) node[below] {$\Step{b}$};


\draw [thick] (6,0) -- (7.5,0);
\draw [thick] (6,0) -- (6,6);
\draw [thick] (7.5,0) -- (7.5,6);
\draw [thick] (6,6) -- (7.5,6);
\draw (6.75,0) node[below] {$\Romannumcolor{3}$};
\draw (6,3) node[left] {128};
\draw (6.75,6) node[above] {4};

\draw[->, ultra thick] (7.5,5) -- (12.5,5);

\draw [thick] (9,6) -- (9.5,6);
\draw [thick] (9.5,6) -- (9.5,0);
\draw [thick] (9.5,0) -- (9,0);
\draw [thick] (9,0) -- (9,6);
\draw (9.25,0) node[below] {$\Romannumcolor{4}$};
\draw (9,3) node[left] {128};
\draw (9.25,6) node[above] {1};

\draw[->, ultra thick] (9.5,2) -- (12.5,2);
\draw (11,5) node[below] {$\Step{c}$};
\draw (11,2) node[below] {$\Step{c}$};

\draw [thick] (12.5,6) -- (15.5,6);
\draw [thick] (15.5,6) -- (15.5,0);
\draw [thick] (15.5,0) -- (12.5,0);
\draw [thick] (12.5,0) -- (12.5,6);
\draw (14,0) node[below] {$\Romannumcolor{5}$};

\draw (12.5,3) node[left] {128};
\draw (14,6) node[above] {16};

\end{tikzpicture}
}
\end{flushright}
\caption{Illustration of the encoding steps (Legend: cf. table \ref{tab:encoding}).}
\label{fig:encoding}
\end{figure}
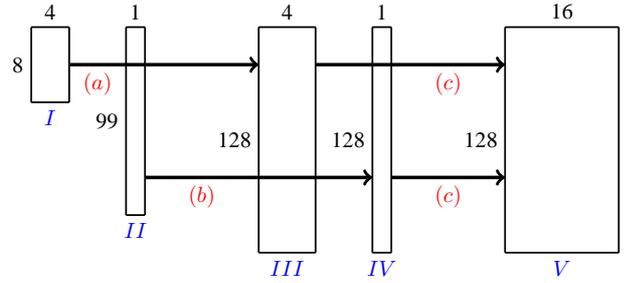

The inner code $\frb^{(1)}$ is partitioned into $16$ disjoint subcodes $\frb^{(2)}_{i}$ with parameters $(16,1,16)$, e.g. $\frb^{(2)}_{0000}$ can be the repetition code of length $16$ and all other elements of the partition are its distinct cosets. The enumeration $i \in \{0000, \dots, 1111\}$ is then protected by four $\frr(1,7)$ codes, one for each bit. Since the subcodes $\frb^{(2)}_{i}$ contain exactly two elements each, we can again partition them into subcodes containing only one element, $\frb^{(2)}_{i,0}$ and $\frb^{(2)}_{i,1}$. The enumeration $\{0,1\}$ is then protected by a $\frr(4,7)$ code. The partition tree is illustrated in Figure~\ref{fig:partition_tree}. Thus, we can encode $4 \cdot 8 + 99 = 131 \geq 128$ bits. Encoding is illustrated in Figure~\ref{fig:encoding} and explained in Table \ref{tab:encoding}.

\begin{table}[h]
\renewcommand{\arraystretch}{1.5}
\centering
\begin{tabular}{p{1cm}|p{6.9cm}}
Block/Step		& Description \\
\hline
\hline
$\Romannumcolor{1}$	& $(8\times4)$-matrix containing $32$ information bits\footnotemark.\\
\hline
$\Step{a}$			& Column-wise encoding in $\fra^{(1)} = \frr(1,7) = \mathcal{C}(128,8,64)$.\\
\hline
$\Romannumcolor{3}$	& Result of Step $\Step{a}$. $(128\times4)$-matrix, whose columns are codwords of $\fra^{(1)}$. Each row provides first partition index $i$ (4 bits) for encoding Step $\Step{c}$ (chooses $\frb^{(2)}_i$).\\
\hline
\hline
$\Romannumcolor{2}$	& $(99\times1)$-matrix containing $99$ information bits\addtocounter{footnote}{-1}\footnotemark.\\
\hline
$\Step{b}$			& Encoding in $\fra^{(2)} = \frr(4,7) = \mathcal{C}(128,99,8)$.\\
\hline
$\Romannumcolor{4}$	& Result of Step $\Step{b}$. $(128\times1)$-matrix which is codword of $\fra^{(2)}$. Each row provides second partition index $j$ (1 bit) for encoding Step $\Step{c}$ (chooses $\frb^{(3)}_{i,j}$).\\
\hline
\hline
$\Step{c}$			& Takes row-wise partition indices $i$ (from $\Romannumcolor{3}$) and $j$ (from $\Romannumcolor{4}$) and writes the codeword contained by $\frb_{i,j}^{(3)}$ (note that this code contains only one codeword) in the corresponding row of $\Romannumcolor{5}$.\\
\hline
$\Romannumcolor{5}$	& GC codeword (length $n=128\cdot16=2048$) obtained by encoding the $k=131$ bits from $\Romannumcolor{1}$ and $\Romannumcolor{2}$.\\
\hline
\hline
\end{tabular}
\vspace{2ex}
\caption{Legend to Figure~\ref{fig:encoding} (Encoding).}
\label{tab:encoding}
\end{table}
\footnotetext{Note that the resulting code has dimension $131 = 32+99$, where $32=4\cdot8$ bits are encoded using $\fra^{(1)}$ and $99$ bits are encoded using $\fra^{(2)}$.}

\begin{figure}[h]
\begin{flushright}
\resizebox{0.45\textwidth}{!}{
\begin{tikzpicture}[scale=0.6]

\draw [thick] (0,6) -- (3,6);
\draw [thick] (3,6) -- (3,0);
\draw [thick] (3,0) -- (0,0);
\draw [thick] (0,0) -- (0,6);
\draw (1.5,0) node[below] {$\Romannumcolor{1}$};

\draw (0,3) node[left] {128};
\draw (1.5,6) node[above] {16};

\draw[->, ultra thick] (3,5) -- (6,5);
\draw (4.5,5) node[below] {$\Step{a}$};

\draw [thick] (6,0) -- (9,0);
\draw [thick] (6,0) -- (6,6);
\draw [thick] (9,0) -- (9,6);
\draw [thick] (6,6) -- (9,6);
\draw (7.5,0) node[below] {$\Romannumcolor{2}$};

\draw (6,3) node[left] {128};
\draw (7.5,6) node[above] {16};

\draw[->, ultra thick] (9,5) -- (12,5);
\draw (10.5,5) node[below] {$\Step{b}$};

\draw [thick] (12,6) -- (13.5,6);
\draw [thick] (13.5,0) -- (13.5,6);
\draw [thick] (12,0) -- (13.5,0);
\draw [thick] (12,0) -- (12,6);
\draw (12.5,0) node[below] {$\Romannumcolor{3}$};

\draw (12,3) node[left] {128};
\draw (12.75,6) node[above] {4};

\draw[->, ultra thick] (13.5,5) -- (16.5,5);
\draw (15,5) node[below] {$\Step{c}$};

\draw [thick] (16.5,6) -- (18,6);
\draw [thick] (18,6) -- (18,0);
\draw [thick] (18,0) -- (16.5,0);
\draw [thick] (16.5,0) -- (16.5,6);
\draw (17,0) node[below] {$\Romannumcolor{4}$};

\draw (16.5,3) node[left] {128};
\draw (17.25,6) node[above] {4};

\draw[->, ultra thick] (18,5) -- (21,5);
\draw (19.5,5) node[below] {$\Step{d}$};

\draw [ultra thick] (17.5,0) -- (17.5,-2);
\draw [ultra thick] (17.5,-2) -- (5.25,-2);
\draw [->,ultra thick] (5.25,-2) -- (5.25,-4.25);

\draw [ultra thick] (2,0) -- (2,-5);
\draw [->, ultra thick] (2,-5) -- (4.75,-5);
\draw (6,-5) node[left] {$\Step{e}$};

\draw [thick] (6,-4) -- (9,-4);
\draw [thick] (9,-4) -- (9,-10);
\draw [thick] (9,-10) -- (6,-10);
\draw [thick] (6,-10) -- (6,-4);
\draw (7.5,-10) node[below] {$\Romannumcolor{5}$};

\draw (6,-7) node[left] {128};
\draw (7.5,-4) node[above] {16};

\draw [->, ultra thick] (9,-5) -- (12,-5);
\draw (10.5,-5) node[below] {$\Step{f}$};

\draw [thick] (12,-4) -- (12.5,-4);
\draw [thick] (12.5,-4) -- (12.5,-10);
\draw [thick] (12.5,-10) -- (12,-10);
\draw [thick] (12,-10) -- (12,-4);
\draw (12.25,-10) node[below] {$\Romannumcolor{6}$};

\draw (12,-7) node[left] {128};
\draw (12.25,-4) node[above] {1};

\draw [->, ultra thick] (12.5,-5) -- (16.5,-5);
\draw (14.5,-5) node[below] {$\Step{g}$};

\draw [thick] (16.5,-4) -- (17,-4);
\draw [thick] (17,-4) -- (17,-10);
\draw [thick] (17,-10) -- (16.5,-10);
\draw [thick] (16.5,-10) -- (16.5,-4);
\draw (16.75,-10) node[below] {$\Romannumcolor{7}$};

\draw (16.5,-7) node[left] {128};
\draw (16.75,-4) node[above] {1};

\draw [->, ultra thick] (17,-5) -- (21,-5);
\draw (19,-5) node[below] {$\Step{h}$};

\end{tikzpicture}
}
\end{flushright}
\caption{Illustration of the decoding steps (Legend: cf. table \ref{tab:decoding}).}
\label{fig:decoding}
\end{figure}
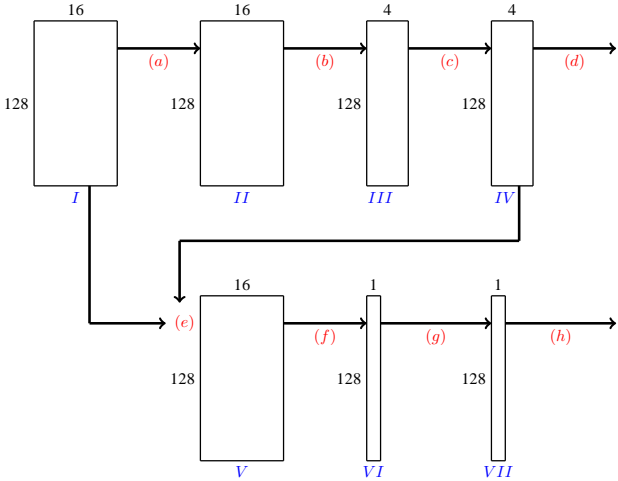

A detailed description of the decoding process is visualized in Figure~\ref{fig:decoding} and explained in Table \ref{tab:decoding}.
The soft information mentioned in Steps $(c)$ and $(g)$ is obtained from the number of errors corrected in Steps $(a)$ and $(e)$ respectively.

\subsection{Analysis of the RM-Example}

We derive an upper bound on the block error probability $\Prob_{err}$ of the code described in Section~\ref{subsec_GMD}. For GC codes, decoding is realized in several steps. We look at the events $S_1, \dots, S_r$, where $S_i$ is the event that decoding in step $i$ fails. Since the decoder is only successful if all steps work properly, we can give an upper bound for $\Prob_{err}$ using the Union Bound:
\begin{align}
\Prob_{err} = \PR{\bigcup_{i=1}^{r} S_i} \leq \sum\limits_{i=1}^{r} \PR{S_i} \label{eq:upper_bound_perr}
\end{align}
Hence, the block error probability is upper-bounded by the sum of the error probabilities of each step.
In the example from Section~\ref{subsec_GMD}, we can group the decoding process into two major steps, namely $S_1$ consisting of steps $\Step{a}-\Step{d}$ and $S_2$ with $\Step{e}-\Step{h}$ (cf. Table \ref{tab:decoding}).
$\PR{S_1}$ can be calculated by transforming the BSC with $p=0.14$ into a \emph{binary error and erasure channel} by ML decoding of the inner Simplex $(16,5,8)$ code. By simulation, we obtain the following parameters of this transformed channel:
\begin{align*}
\PR{\text{error}} &= 0.020698, \\
\PR{\text{erasure}} &= 0.155532.
\end{align*}
The error-erasure decoder of the outer $\frr(1,7) = \C(128,8,64)$ code can decode correctly, if $2\tau+\delta<64$, where $\tau$ is the number of errors and $\delta$ is the number of erasures.
Using this condition, we obtain
\begin{align}
\PR{S_1} &= \PR{2\tau+\delta \geq 64} \notag \\
         &= \sum\limits_{i=0}^{128}\PR{\delta=i}\PR{2\tau\geq 64-i \, | \, \delta=i} \notag \\
		 &\approx 9.51 \cdot 10^{-12} \label{eq:PS1}
\end{align}
The probability $\PR{S_2}$ can be calculated similarly. It turns out that $\PR{S_2} \approx 1.48 \cdot 10^{-9}$.
Using these results, we obtain the following upper bound on $\Prob_{err}$:
\begin{align*}
\Prob_{err} \leq 9.51 \cdot 10^{-12} + 1.48\cdot 10^{-9} \approx 1.49 \cdot 10^{-9} 
\end{align*}
However, this probability can be further decreased by using GMD decoding. This effect is not easy to analyze analytically, but has a large impact on the error probability. Simulations have shown that the actual block error probability is given by:
\begin{align}
\Prob_{err} \approx 5.37 \cdot 10^{-10} \label{eq:RM_perr_result}
\end{align}
Compared to the code construction used in \cite{MaesCryptoPaper2012}, we obtain a smaller block error probability using GC codes. We have also decreased the codeword length from $2226$ to $2048$. Another advantage of our construction is that decoding is easier to implement, since we only use codes with decoders working over $\FF_2$. Table~\ref{tab:summary} summarizes the improvements.

\begin{table}[h]
\renewcommand{\arraystretch}{1.5}
\centering
\begin{tabular}{p{1cm}|p{6.9cm}}
Block/Step		& Description \\
\hline
\hline
$\Romannumcolor{1}$	& $(128\times16)$-matrix containing the received word. Rows are codewords of $\frb^{(1)}=\frr(1,4)$ (Simplex code) plus error. \\
\hline
\hline
$\Step{a}$			& Row-wise ML decoding in $\frb^{(1)}$. Result: $c \in \frb^{(1)}$ or $\otimes^{16}$ (erasure if closest codeword not unique)\\
\hline
$\Romannumcolor{2}$	& Result of Step $\Step{a}$. Rows are codewords of $\frb^{(1)}$ or $\otimes^{16}$\\
\hline
\hline
$\Step{b}$			& Remapping of every row (codewords of $\frb^{(1)}$) to index (4 bits) of the partition which contains the row. If erasure, result: $\otimes^{4}$.\\
\hline
$\Romannumcolor{3}$	& Result of Step $\Step{b}$. Rows are $\in \{0,1\}^4 \cup \{\otimes^{4}\}$.\\
\hline
\hline
$\Step{c}$			& Column-wise error-erasure decoding (optional: GMD using soft information obtained from step $\Step{a}$) in $\fra^{(1)} = \frr(1,7) = \mathcal{C}(128,8,64)$. If decoding fails: Declare failure of algorithm.\\
\hline
$\Romannumcolor{4}$	& Result of Step $\Step{c}$. Columns are $\in \frr(1,7)$. Rows give indices $i$ (4 bits) which specify in which partition $\frb_i^{(2)}$ the rows must be decoded in the second part of the algorithm. \\
\hline
\hline
$\Step{d}$			& Extraction of the first $32 = 4 \cdot 8$ information bits (each column of $\Romannumcolor{4}$ is a codeword of a $\mathcal{C}(128,8,64)$ code which corresponds to exactly one information word of length $8$).\\
\hline
\hline
$\Step{e}$			& Row-wise ML decoding in $\frb_i^{(2)}$, where $i$ denotes the partition index for each row given by the corresponding row of $\Romannumcolor{4}$.\\
\hline
$\Romannumcolor{5}$	& Result of Step $\Step{e}$. Rows are codewords of $\frb_i^{(2)}$ or $\otimes^{16}$ (if closest codeword not unique).\\
\hline
\hline
$\Step{f}$			& Remapping of every row (codewords of $\frb_i^{(2)}$) to index $j$ (1 bit) of the partition $\frb_{i,j}^{(3)}$ of $\frb_i^{(2)}$ which contains the row. If erasure: $\otimes$.\\
\hline
$\Romannumcolor{6}$	& Result of Step $\Step{f}$.\\
\hline
\hline
$\Step{g}$			& Error-erasure decoding (optional: GMD using soft information obtained from step $\Step{e}$) of the column in $\fra^{(2)} = \frr(4,7) = \mathcal{C}(128,99,8)$. If decoding fails: Declare failure of algorithm.  \\
\hline
$\Romannumcolor{7}$	& Result of Step $\Step{g}$. Column contains codeword of $\fra^{(2)}$.\\
\hline
\hline
$\Step{h}$			& Extraction of remaining $99$ information bits which correspond to the $\fra^{(2)} = \mathcal{C}(128,99,8)$ codeword in Block $\Romannumcolor{7}$.\\
\hline
\hline
\end{tabular}
\vspace{2ex}
\caption{Legend to Figure \ref{fig:decoding} (Decoding).}
\label{tab:decoding}
\end{table}

\subsection{Reed-Solomon Construction}
\label{subsec:RSEx}

The result of Section~\ref{subsec:RMEx} provides a code construction which improves commonly used coding schemes for key reproduction in PUFs.
However, the use of RM codes as outer codes in the GC scheme is not as flexible as sometimes desired.
Thus, we show how GC codes can be constructed using RS codes as outer codes and improve the code construction from Section~\ref{subsec:RSEx} in the code length.

The possible length $n_o$ of the outer RS codes is upper bounded by their field size, which again is restricted by the number of partitionings.
This means that the length of the inner codes $n_i$ must be sufficiently large, such that at each partition level $i$, partitions of $\frb^{(i)}$ in more subcodes than the length $n_o$ of the outer codes are possible.

We first illustrate how much the code rate can be reduced when using RS codes in a concatenated scheme by giving an example of an ordinarily concatenated code based on RS codes.
The example uses a $\frr(1,5) = \mathcal{C}(32,6,16)$ code as inner code which transforms the BSC with $p=0.14$ into a binary error and erasure channel with $\PR{\text{error}} = 0.003170$ and  $\PR{\text{erasure}} = 0.017605$ using ML decoding.
As outer code, we use a $\RSq{2^6;n,k}$ code with $n\leq2^6=64$ and $k = 22$ because the overall dimension of the code must be $6k\geq128$ bits.
If we choose $n=64$, we obtain a code with length $n=64 \cdot 32 = 2048$ and we can calculate that $\Prob_{err} \approx 6.79 \cdot 10^{-37}$ using the same equation as \eqref{eq:PS1}.
Since this probability is by far smaller than necessary, we can use the flexibility of RS codes and reduce $n$ arbitrarily.
This is easy to realize with the same decoder as for the $\RSq{2^6;64,k}$ code by declaring some codeword positions to be erasures.
If we use a $\RSq{2^6;36,22}$ code, the code length can be reduced to $1152$ and we obtain a block error probability of $\Prob_{err} \approx 1.19 \cdot 10^{-10}$.
Note that we have already reduced the code length by half compared to the construction in \cite{MaesCryptoPaper2012}.

\begin{figure}[h]
\tikzstyle{level 1}=[level distance=4cm, sibling distance=10cm]
\tikzstyle{level 2}=[level distance=4cm, sibling distance=10cm]
\tikzstyle{level 3}=[level distance=4cm, sibling distance=6cm]
\tikzstyle{level 4}=[level distance=4cm, sibling distance=1.5cm]

\tikzstyle{bag} = [text width=10em, text centered]
\tikzstyle{end} = [circle, minimum width=3pt,fill, inner sep=0pt]

\centering
\hspace{-1cm}
\resizebox{0.52\textwidth}{!}{
\begin{tikzpicture}[scale=0.4] 
\draw (0,-2) 	node {$\dots$};
\draw (0,-4) 	node {$\dots$};
\draw (0,-10) 	node {$\vdots$};
\draw (-5,-6) 	node {$\dots$};
\draw (-5,-8) 	node {$\dots$};
\draw (5,-6) 	node {$\vdots$};

\draw (12,0) 	node {Level 1};
\draw (12,-4) 	node {Level 2};
\draw (12,-8) 	node {Level 3};
\draw (12,-12)	node {Level 4};
\draw (12,-2) 	node {$\longleftarrow \fra^{(1)}(2^{5};32,2,31)$};
\draw (12,-6) 	node {$\longleftarrow \fra^{(2)}(2^{5};32,19,12)$};
\draw (12,-10)	node {$\longleftarrow \fra^{(3)}(2^1;32,24,4)$};
\node[bag] {$\frb^{(1)}(32,11,12)$}
    child {
        node[bag] {$\frb^{(2)}_{00000}(32,6,16)$}       
            child {
                node[bag] {$\frb^{(3)}_{00000,00000}(32,1,32)$}
             	child {
             	node[bag] {$\frb^{(4)}_{00000,000000,1}$}
                	edge from parent
                	node[left] {$0$}
                }
                child{
                	node[bag] {$\frb^{(4)}_{00000,000000,0}$}
                	edge from parent
                	node[right] {$1$}
             	}
                edge from parent
                node[left] {$00000$}
            }
            child {
                node[bag] {$\frb^{(3)}_{00000,11111}(32,1,32)$}
                 edge from parent
                node[right] {$11111$}
            }
            edge from parent
            node[left] {$00000$}
    }
    child {
        node[bag] {$\frb^{(2)}_{11111}(32,6,16)$}
            edge from parent
            node[right] {$11111$}
    };
\end{tikzpicture}
}
\caption{Partition of the inner extended BCH code $\frb^{(1)}(32,11,12)$.}
\label{fig:partition_tree_rs}
\end{figure}
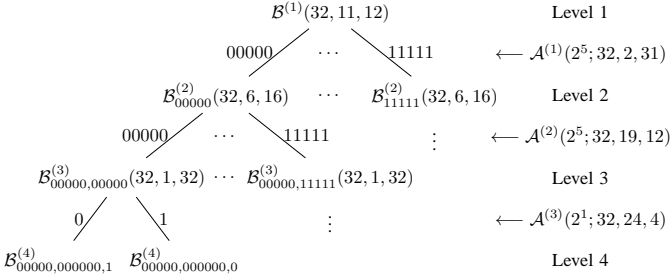

In the following, we give an example that reduces the code length even more by using GC codes.
We partition an extended BCH \cite{Bossert1999} code $\frb^{(1)}(32,11,12)$ into thirty-two $\frb^{(2)}_i(32,6,16)$ codes ($i \in \{0,1\}^5$), which we again partition in $\frb^{(3)}_{i,j}(32,1,32)$ codes ($j \in \{0,1\}^5$).
As outer codes we use RS codes to protect the partitions, e.g. an $\fra^{(1)} = \RSq{2^{5};32,2,31}$ code to protect the partitioning from level~$1$ to level~$2$ and an $\fra^{(2)} = \RSq{2^{5};32,19,12}$ code between levels~$2$ and~$3$.
The partition from level~$3$ to level~$4$ is protected by an $\fra^{(3)} = \frr(32,26,4)$ code.
The partition tree for this example is visualized in Figure~\ref{fig:partition_tree_rs}.
Encoding and decoding is done similarly to the RM example in Section~\ref{subsec:RMEx}.

We analyze the decoding capabilities of the code step by step.
In \textbf{Step 1}, ML decoding in $\frb^{(1)}(32,11,12)$ transforms the channel in a binary error and erasure channel\footnote{The parameters of the transformed channel have been calculated by simulation of ML decoding of the inner code.} with $\PR{\text{error}} = 0.037808$ and  $\PR{\text{erasure}} = 0.174488$. Decoding up to half the minimum distance in the outer code would leave us with $\PR{S_1}\approx 1.03 \cdot 10^{-8}$, which is too high. But since the $\RSq{2^{5};32,2,31}$ code has a low rate, we can apply \emph{Power Decoding} (cf. Section \ref{subsec_rs}) and obtain an error probability of $\PR{S_1}\approx1.48 \cdot 10^{-11}$.

\textbf{Step 2} transforms the BSC into a binary error and erasure channel with $\PR{\text{error}} = 0.0032167$ and  $\PR{\text{erasure}} = 0.0175397$. Hence, decoding in $\RSq{2^{5};32,19,12}$ yields an error probability of $\PR{S_2}\approx3.11 \cdot 10^{-10}$.

The \textbf{last step} has $\PR{S_3}\approx2.13 \cdot 10^{-11}$.
Hence, the overall block error probability is upper bounded by
\begin{equation*}
\Prob_{err} \underset{\eqref{eq:upper_bound_perr}}{\leq} \PR{S_1} + \PR{S_2} + \PR{S_3} \approx 3.47 \cdot 10^{-10} 
\end{equation*}
Thus, the example satisfies the constraints and reduces the code length to $n = 32 \cdot 32 = 1024$.

\section{Evaluation and Conclusion}
\label{sec_conclusion}

We explained how coding theory is used for reproducing cryptographic keys using PUFs.
Furthermore, we proposed code constructions and decoding methods which improve existing coding schemes for PUFs and illustrated these by giving examples.
Table~\ref{tab:summary} summarizes the properties of the example constructions.
It can be seen that our approach can achieve significantly reduced codelengths, block error probabilities or decoding complexity.
In future work, more methods from coding theory can be examined for suitability in this setting.

\section*{Acknowledgment}
The authors thank Vladimir Sidorenko for the valuable discussions.

\begin{table}[h]
\renewcommand{\arraystretch}{1.5}
\centering
\begin{tabular}{p{2cm}|p{2cm}|p{1.3cm}|p{1.7cm}}
Code (Section)
& $\Prob_{err}$
& Length
& Largest Field\footnotemark \\
\hline \hline
BCH Rep. \cite{MaesCryptoPaper2012}
& $10^{-9}$
& $2226$
& $\FF_{2^8}$ (BCH) \\
\hline
GC RM (\ref{subsec:RMEx})
& $5.37 \cdot 10^{-10}$
& $2048$
& $\FF_2$ \\
\hline
RS (\ref{subsec:RSEx})
& $6.79 \cdot 10^{-37}$
& $2048$
& $\FF_{2^6}$ \\
\hline
RS (\ref{subsec:RSEx})
& $1.19 \cdot 10^{-10}$
& $1152$
& $\FF_{2^6}$ \\
\hline
GC RS (\ref{subsec:RSEx})
& $3.47 \cdot 10^{-10}$
& $1024$
& $\FF_{2^5}$ \\
\end{tabular}
\vspace{2ex}
\caption{Comparison between the code constructions.}
\label{tab:summary}
\end{table}
\footnotetext{Largest field used by decoder. Operations over small fields are usually easier to implement.}

\bibliographystyle{IEEEtran}
\bibliography{puf_gcc}

\end{document}